\documentclass[letterpaper]{article}
\usepackage{aaai}
\usepackage{times}
\usepackage{helvet}
\usepackage{courier}
\usepackage{color}
\usepackage[caption=false]{subfig}
\usepackage{booktabs}
\newcommand{\ra}[1]{\renewcommand{\arraystretch}{#1}}
\usepackage{url}
\usepackage{verbatim} 
\usepackage{multirow}
\usepackage{algorithm}
\usepackage[noend]{algorithmic}
\usepackage{xspace}
\usepackage{graphicx}
\usepackage{epsfig}
\usepackage{amsmath}
\usepackage{amssymb}
\usepackage{times}
\usepackage{paralist}
\newcommand{\cut}[1]{}

\newcommand{\punished}{punished\xspace}
\newcommand{\rewarded}{rewarded\xspace}
\newcommand{\punishments}{punishments\xspace}
\newcommand{\rewards}{rewards\xspace}
\newcommand{\punishment}{punishment\xspace}
\newcommand{\reward}{reward\xspace}

\newcommand{\ppp}{p}
\newcommand{\qqq}{q}

\begin{document}

\title{How Community Feedback Shapes User Behavior}
\author{Justin Cheng$^*$, Cristian Danescu-Niculescu-Mizil$^\dagger$, Jure Leskovec$^*$\\
$^*$Stanford University, $^\dagger$Max Planck Institute SWS\\
\tt \footnotesize
jcccf$\vert$jure@cs.stanford.edu, cristian@mpi-sws.org
}
\maketitle
\begin{abstract}
\begin{quote}

Social media systems rely on user feedback and rating mechanisms for personalization, ranking, and content filtering. 
However, when users evaluate content contributed by fellow users (e.g., by liking a post or voting on a comment), these evaluations create complex social feedback effects.
This paper investigates how ratings on a piece of content affect its author's future behavior.
By studying four large comment-based news communities, we find that negative feedback leads to significant behavioral changes that are detrimental to the community.
Not only do authors of negatively-evaluated content contribute more, but also their future posts are of lower quality, and are perceived by the community as such.
Moreover, these authors are more likely to subsequently evaluate their fellow users negatively, percolating these effects through the community.
In contrast, positive feedback does not carry similar effects, and neither encourages rewarded authors to write more, nor improves the quality of their posts. 
Interestingly, the authors that receive no feedback are most likely to leave a community.
Furthermore, a structural analysis of the voter network reveals that evaluations polarize the community the most when positive and negative votes are equally split.

\end{quote}
\end{abstract}

\section{Introduction}
\label{sec:intro}

The ability of users to rate content and provide feedback is a defining characteristic of today's social media systems.
These ratings enable the discovery of high-quality and trending content, as well as personalized content ranking, filtering, and recommendation.
However, when these ratings apply to content generated by fellow users---helpfulness ratings of product reviews, likes on Facebook posts, or up-votes on news comments or forum posts---evaluations also become a mean of social interaction.
This can create social feedback loops that affect the behavior of the author whose content was evaluated, as well as the entire community.

Online rating and evaluation systems have been extensively researched in the past. The focus has primarily been on predicting user-generated ratings \cite{kim2006automatically,Ghose:ProceedingsOfTheInternationalConferenceOnElectronic:2007,otterbacher2009helpfulness,anderson12status} and on understanding their effects at the community-level 
\cite{DanescuNiculescuMizil+al:09a,Anderson:ProceedingsOfKdd:2012,Muchnik:Science:2013,Sipos:Www:2014}.
However, little attention has been dedicated to studying the effects that ratings have on the behavior of the author whose content is being evaluated.

Ideally, feedback would lead users to behave in ways that benefit the community.
Indeed, if positive ratings act as ``reward'' stimuli and negative ratings act as ``punishment'' stimuli, the {\em operant conditioning} framework from behavioral psychology \cite{Skinner:TheBehaviorOfOrganismsAnExperimentalAnalysis:1938} predicts that community feedback should guide authors to generate better content in the future, and that \punished authors will contribute less than \rewarded authors.
However, despite being one of the fundamental frameworks in behavioral psychology,
there is limited empirical evidence of operant conditioning effects on humans\footnote{The framework was developed and tested mainly through experiments on animal behavior (e.g., rats and pigeons); the lack of human experimentation can be attributed to methodological and ethical issues, especially with regards to punishment stimuli (e.g. electric shocks).} \cite{Baron:BehavAnal:1991}.
Moreover, it remains unclear whether community feedback in complex online social systems brings the intuitive beneficial effects predicted by this theory.

In this work we develop a methodology for quantifying and comparing the effects of \rewards and \punishments on multiple facets of the author's future behavior in the community, and relate these 
effects
 to the broader theoretical framework of {\em operant conditioning}.
In particular, we seek to understand whether community feedback regulates the quality and quantity of 
a
user's future contributions in a way that benefits the community.

By applying our methodology to four large online news communities for which we have complete article commenting and comment voting data (about 140 million votes on 42 million comments), we discover that community feedback does {\em not} appear to drive the behavior of users in a direction that is beneficial to the community, as predicted by the operant conditioning framework. 
Instead, we find that community feedback is likely to perpetuate
undesired behavior.
In particular, \punished authors actually write worse\footnote{One important subtlety here is that the observed quality of a post (i.e., the proportion of up-votes) is not entirely a direct consequence of the actual textual quality of the post, but is also affected by community bias effects. We account for this through experiments specifically designed to disentangle these two factors.} in subsequent posts, while \rewarded authors do not improve significantly.

In spite of this detrimental effect on content {\em quality}, it is conceivable that community feedback still helps regulate {\em quantity} by selectively discouraging contributions from \punished authors and encouraging \rewarded authors to contribute more.  Surprisingly, we find that negative feedback actually leads to more (and more frequent) future contributions than positive feedback does.\footnote{We note that these observations cannot simply be attributed to flame wars, as they spread over a much larger time scale.}
Taken together, our findings suggest that the content evaluation mechanisms currently implemented in social media systems have effects contrary to the interest of the community.

To further understand differences in social mechanisms causing these behavior changes, we conducted a structural analysis of the voter network around popular posts. We discover that not only does positive and negative feedback tend to come from communities of users, but that the voting network is most polarized when votes are split equally between up- and down-votes.

These observations underscore the asymmetry between the effects of positive and negative feedback: the detrimental impact of \punishments is much more noticeable than the beneficial impact of \rewards. This asymmetry echoes the {\em negativity effect} studied extensively in social psychology literature: negative events have a greater impact on individuals than positive events of the same intensity~\cite{Kanouse:NegativityInEvaluations:1972,baumeister2001bad}.

To summarize our contributions, in this paper we
\begin{itemize}
\item validate through a crowdsourcing experiment that the proportion of up-votes is a robust metric for measuring and aggregating community feedback,
\item introduce a framework based on propensity score matching for quantifying the effects of community feedback on a user's post quality,
\item discover that effects of community evaluations are generally detrimental to the community, contradicting the intuition brought up by the operant conditioning theory, and
\item reveal an important asymmetry between the mechanisms underlying negative and positive feedback.
\end{itemize}

Our results lead to a better understanding of how users react to peer evaluations, and point to ways in which online rating mechanisms can be improved to better serve individuals, as well as entire communities.

\section{Further Related Work}
\label{sec:related}
Our contributions come in the context of an extensive literature examining social media voting systems.
One major research direction is concerned with predicting the helpfulness ratings of product reviews starting from textual and social factors \cite{Ghose:ProceedingsOfTheInternationalConferenceOnElectronic:2007,liu2007low,otterbacher2009helpfulness,tsur2009revrank,Lu:2010:ESC:1772690.1772761,Mudambi:MisQuarterly:2010} and understanding the underlaying social dynamics \cite{chen2008,DanescuNiculescuMizil+al:09a,wu2010opinion,Sipos:Www:2014}.
The mechanisms driving user voting behavior and the related community effects have been studied in other contexts, such as Q\&A sites \cite{Anderson:ProceedingsOfKdd:2012}, Wikipedia \cite{burke08wikipedia,Leskovec+Huttenlocher+Kleinberg:2010a,anderson12status}, YouTube \cite{Siersdorfer:ProceedingsOfWww:2010}, social news aggregation sites \cite{Lampe:2004:SBD:985692.985761,Lampe:2005:FDE:1099203.1099206,Muchnik:Science:2013} and online multiplayer games \cite{Shores:Cscw:2014}.
Our work adds an important dimension to this general line of research, by providing a framework for analyzing the effects votes have on the author of the evaluated content.

The setting considered in this paper, that of comments on news sites and blogs, has also been used to study other social phenomena such as controversy \cite{Chen:JournalOfConsumerResearchJournalOfConsumer:2013}, political polarization \cite{Park:ProceedingsOfCscw:2011,Balasubramanyan:ProceedingsOfIcwsm:2012}, and community formation \cite{Gomez+Kaltenbrunner+Lopez:2008a,gonzalez2010structure}.
News commenting systems have also been analyzed from a community design perspective \cite{Mishne:ThirdAnnualWorkshopOnTheWebloggingEcosystem:2006,Gilbert:HawaiiInternationalConferenceOnSystemSciences:2009,Diakopoulos:2011:TQD:1958824.1958844}, including a particular focus on understanding what types of articles are likely to attract a large volume of user comments \cite{Tsagkias:2009:PVC:1645953.1646225,Yano:ProcOfIcwsm:2010}.
In contrast, our analysis focuses on the effects of voting on the behavior of the author whose content is being evaluated. 

Our findings here reveal that negative feedback does not lead to a decrease of undesired user behavior, but rather attenuates it.
Given the difficulty of moderating undesired user behavior, it is worth pointing out that anti-social behavior in social media systems is a growing concern \cite{Heymann:InternetComputingIeee:2007}, as emphasized by work on review spamming \cite{Lim:2010:DPR:1871437.1871557,mukherjeespotting,Ott:Www2012:2012}, trolling \cite{shachaf2010beyond}, social deviance \cite{Shores:Cscw:2014} and online harassment \cite{yin2009detection}.

\section{Measuring Community Feedback}
\label{sec:res1}

\begin{table*}[!ht]
\small
\centering
\ra{1.3}
\begin{tabular}{lllllll}\toprule
    \multirow{2}{*}{\textbf{Community}} & \multirow{2}{*}{\textbf{\# Threads}} & \multirow{2}{*}{\textbf{\# Posts}} & \multirow{2}{*}{\textbf{\# Votes (Prop. Up-votes)}} & \multirow{2}{*}{\textbf{\# Registered Users}} & \multicolumn{2}{l}{\textbf{Prop. Up-votes}} \\
    & & & & & $Q_1$ & $Q_3$\\
    \hline
    CNN & 200,576 & 26,552,104 & 58,088,478 (0.82) & 1,111,755 & 0.73 & 1.00\\
    IGN & 682,870 & 7,967,414 & 40,302,961 (0.84) & 289,576 & 0.69 & 1.00\\
    Breitbart & 376,526 & 4,376,369 & 18,559,688 (0.94) & 214,129 & 0.96 & 1.00 \\
    allkpop & 35,620 & 3,901,487 & 20,306,076 (0.95) & 198,922 & 0.84 & 1.00 \\
\bottomrule
\end{tabular}
\caption{Summary statistics of the four communities analyzed in this study. The lower ($Q_1$) and upper ($Q_3$) quartiles for the proportion of up-votes only takes into account posts with at least ten votes.}
\label{tab:communities}
\end{table*}

We aim to develop a methodology for studying the subtle effects of community-provided feedback on the behavior of content authors in realistic large-scale settings.  To this end, we start by describing a longitudinal dataset where millions of users explicitly evaluate each others' content.
Following that, we discuss a crowdsourcing experiment that helps establish a robust aggregate measure of community feedback.

\subsection{Dataset description}

We investigate four online news communities: \emph{CNN.com} (general news), \emph{Breitbart.com} (political news), \emph{IGN.com} (computer gaming), and \emph{Allkpop.com} (Korean entertainment), selected based on diversity and their large size.
Common to all these sites is that community members post comments on (news) articles, where
each comment can then be up- or down-voted by other users.
We refer to a comment as a {\em post} and to all posts relating to the same article as a {\em thread}.

From the commenting service provider, we obtained complete timestamped trace of user activity from March 2012 to August 2013.\footnote{This is prior to an interface change that hides down-votes.} 
We restrict our analysis to users who joined a given community after March 2012, so that we are able to track users' behavior from their ``birth'' onwards. As shown in Table~\ref{tab:communities} the data includes 1.2 million threads with 42 million comments, and 140 million votes from 1.8 million different users. In all communities around 50\% of posts receive at least one vote, and 10\% receive at least 10 votes.

\subsection{Measures of Community Feedback}

Given a post with with some number of up- and down-votes we next require a measure that aggregates the post's votes into a single number, and that corresponds to the magnitude of reward/punishment received by the author of the post.
However, it is not a priori clear how to design such a measure.
For example, consider a post that received $P$ up-votes and $N$ down-votes.
How can we combine $P$ and $N$ into a single number that best reflects the overall evaluation of the community?
There are several natural candidates for such a measure: the total number of up-votes ($P$) received by a post, the proportion of up-votes ($P/(P+N)$), or the difference in number of up-/down-votes ($P-N$).
However, each of these measures has a particular drawback: $P$ does not consider the number of down-votes (e.g., 10+/0- vs. 10+/20-); $P/(P+N)$ does not differentiate between absolute numbers of votes received (e.g., 4+/1- vs. 40+/10-); and, $P-N$ does not consider the effect relative to the total number of posts (e.g., 5+/0- vs. 50+/45-).

To understand what a person's ``utility function'' for votes is, we conducted an Amazon Mechanical Turk experiment that asked users how they would perceive receiving a given number of up- and down-votes. On a seven-point Likert scale, workers rated how they would feel about receiving a certain number of up- and down-votes on a comment that they made. The number of up- and down-votes was varied between 0 and 20, and each worker responded to 10 randomly-selected pairs of up-votes and down-votes. We then took the average response as the mean rating for each pair.
66 workers labeled 4,302 pairs in total, with each pair obtaining at least 9 independent evaluations.

\begin{figure}[t]
  \centering
  \includegraphics[width=\linewidth]{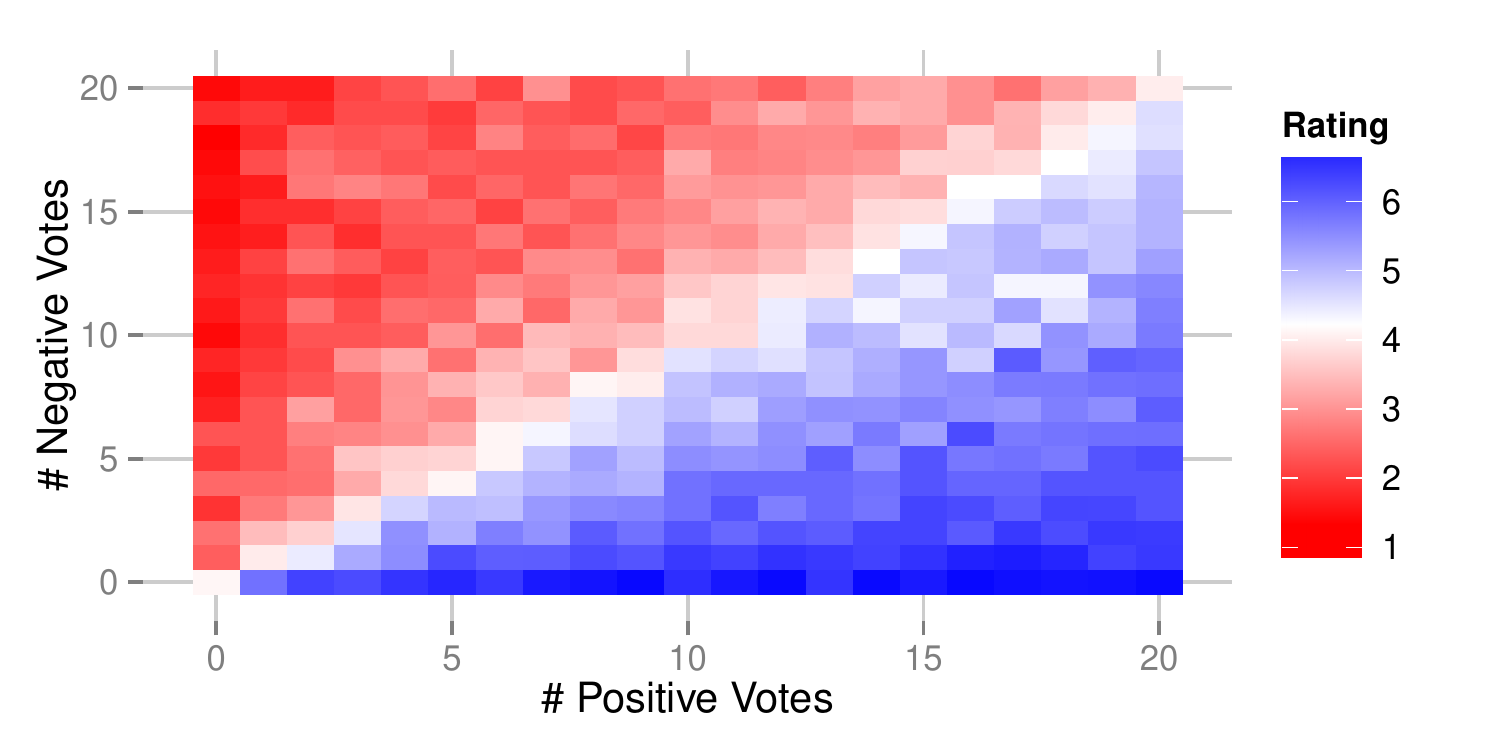}
  \caption{People perceive votes received as proportions, rather than as absolute numbers. Higher ratings correspond to more positive perceptions.}
  \label{fig:voting_grid}
\end{figure}

We find that the proportion of up-votes ($P/(P+N)$) is a very good measure of how positively a user perceives a certain number of up-votes and down-votes.
In Figure \ref{fig:voting_grid}, we notice a strong ``diagonal'' effect, suggesting that increasing the total number of votes received, while maintaining the proportion of up-votes constant, does not significantly alter a user's perception.
In Table \ref{tab:measure} we evaluate how different measures correlate with human ratings, and find that $P/(P+N)$ explains almost all the variance and achieves the highest $R^2$ of 0.92.
While other more complex measures could result in slightly higher $R^2$, we subsequently use $p=P/(P+N)$ because it is both intuitive and fairly robust.

\begin{table}[b]
\small
\centering
\ra{1.3}
\begin{tabular}{llll}\toprule
    \textbf{Measure} & \textbf{$\mathbf{R^2}$} & \textbf{F-Statistic} & \textbf{$\mathbf{p}$-value} \\
    \hline
    $P$ & 0.410 & $F(439) = 306.1$ & $<10^{-16}$ \\
    $P - N$ & 0.879 & $F(438) = 1603$ & $< 10^{-16}$ \\
    $P / (P + N)$ & \textbf{0.920} & $F(438) = 5012$ & $<10^{-16}$ \\
\bottomrule
\end{tabular}
\caption{The proportion of up-votes $p=P/(P+N)$ best captures a person's perception of up-voting and down-voting, according to a crowdsourcing experiment.}
\label{tab:measure}
\end{table}

Thus, for the rest of the paper we use the {\em proportion of up-votes} (denoted as $p$) as the measure of the overall feedback of the community. We consider a post to be {\em positively evaluated} if the proportion of up-votes received is in the upper quartile $Q_3$ (75\textsuperscript{th} percentile) of all posts, and {\em negatively evaluated} if the fraction is instead in the lower quartile $Q_1$ (25\textsuperscript{th} percentile).
This lets us account for differences in community voting norms: in some communities, a post may be perceived as bad, even with a high fraction of up-votes (e.g. Breitbart).
As the proportion of up-votes is skewed in most communities, at the 75th percentile all votes already tend to be up-votes (i.e, feedback is 100\% positive).
Further, in order to obtain sufficient precision of community feedback, we require that these posts have at least ten votes.

Unless specified otherwise, all reported observations are consistent across all four communities we studied. For brevity, the figures that follow are reported only for CNN, with error bars indicating $95\%$ confidence intervals.

\section{Post Quality}
\label{sec:quality}

The {\em operant conditioning framework} posits that an individual's behavior is guided by the consequences of its past behavior~\cite{Skinner:TheBehaviorOfOrganismsAnExperimentalAnalysis:1938}.  
In our setting, this would predict that community feedback would lead users to produce better content.
Specifically, we expect that users \punished via negative feedback would either improve the quality of their posts, or contribute less.
Similarly, users \rewarded by receiving positive feedback would write higher quality posts, and contribute more often.

\begin{figure}[t]
  \centering
  \includegraphics[width=0.9\linewidth]{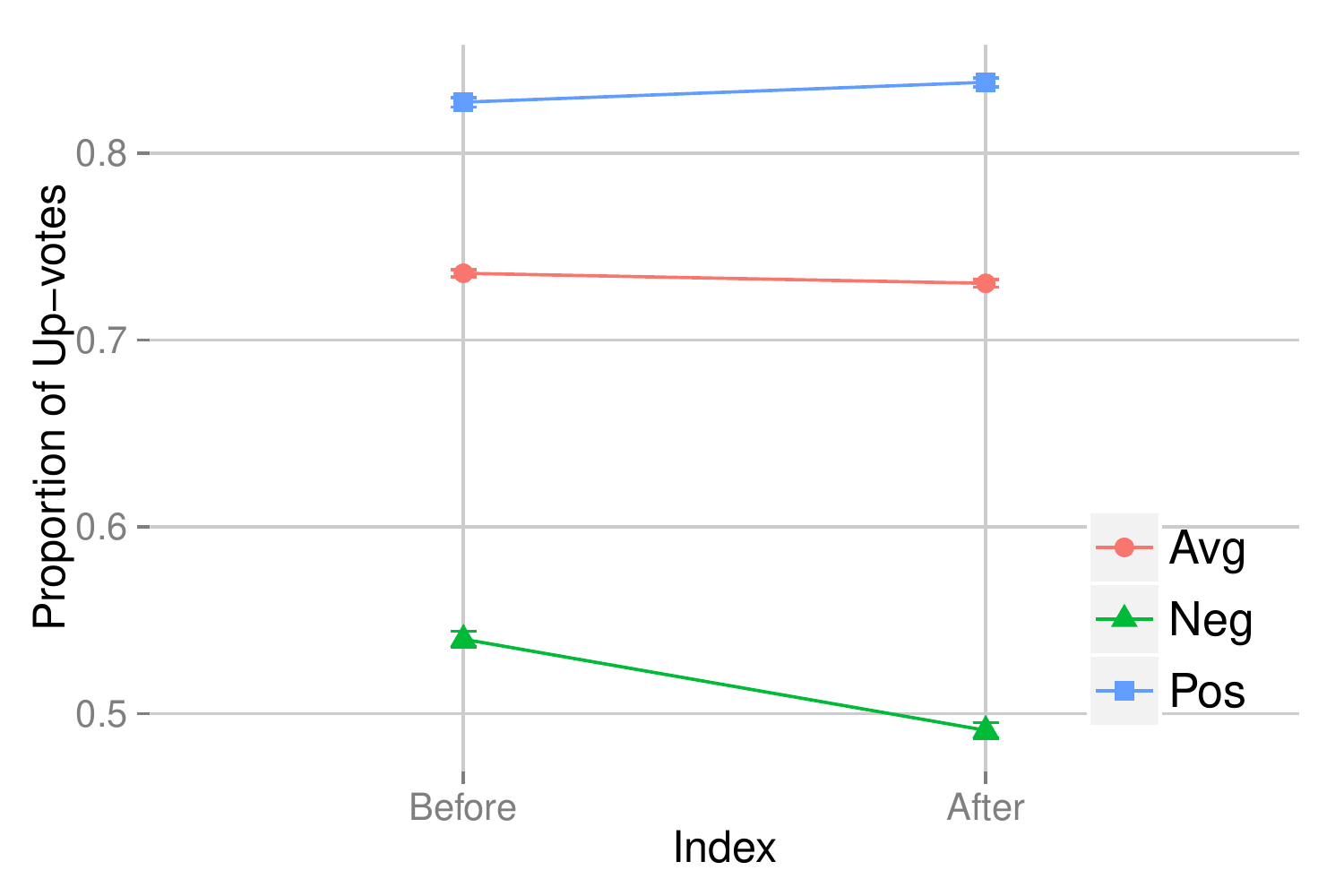}
  \caption{Proportion of up-votes before/after a user receives a positive (``Pos''), negative (``Neg'') or neutral (``Avg'') evaluation. After a positive evaluation, future evaluations of an author's posts do not differ significantly from before. However, after a negative evaluation, an author receives worse evaluations than before.}
  \label{fig:orig_base}
  \vspace{-0.5em}
\end{figure}

In this section, we focus on understanding the effects of positive and negative feedback on the the quality of one's posts. 
We start by simply measuring the {\em post quality} as the {\em proportion of up-votes} a given post received, denoted by $\ppp$.
Figure~\ref{fig:orig_base} plots the proportion of up-votes $\ppp$ as a function of time for users who received a positive, negative or neutral evaluation.
We compare the proportion of up-votes of posts written before receiving the positive/negative evaluation with that of posts written after the evaluation. Interestingly, there is no significant difference for positively evaluated users (i.e., there is no significant difference between $\ppp$ before and after the evaluation event).

In the case of a negative evaluation however, \punishment leads to worse community feedback in the future.
More precisely, the difference in the proportion of up-votes received by a user before/after the feedback event is statistically significant at $p<0.05$.
This means that negative feedback seems to have exactly the opposite effect than predicted by the operant conditioning framework~\cite{Skinner:TheBehaviorOfOrganismsAnExperimentalAnalysis:1938}.
Rather than feedback leading to better posts, Figure~\ref{fig:orig_base} suggests that \punished users actually get worse, not better, after receiving a negative evaluation.

\subsection{Textual vs. Community Effects}
One important subtlety is that the observed proportion of up-votes is not entirely a direct consequence of the actual textual quality of the post, but could also be due to a community's biased perception of a user.
In particular, the drop in the proportion of up-votes received by users after negative evaluations observed in Figure \ref{fig:orig_base} could be explained by two non-mutually exclusive phenomena: 
(1) after negative evaluations, the user writes posts that are of lower quality than before (\emph{textual quality}), or 
(2) users that are known to produce low quality posts automatically receive lower evaluations in the future, regardless of the actual textual quality of the post (\emph{community bias}).

We disentangle these two effects through
a methodology inspired by propensity matching, a statistical technique used to support causality claims in observational studies \cite{ROSENBAUM:Biometrika:1983}.

First, we build a machine learning model that predicts a post's quality ($\qqq$) by training a binomial regression model using {\em only} textual features extracted from the post's content, i.e. $\qqq$ is the {\em predicted} proportion of a post's up-votes.
This way we are able to model the relationship between the content of a post and the post's {\em quality}.
This model was trained on half the posts in the community, and used to predict $\qqq$ for the other half (mean $R = 0.22$).

We validate this model using human-labeled text quality scores obtained for a sample of posts ($n=171$).
Using Crowdflower, a crowdsourcing platform, workers were asked to label posts as either ``good'' (defined as something that a user would want to read, or that contributes to the discussion), or ``bad'' (the opposite).
They were only shown the text of individual posts, and no information about the post's author.
Ten workers independently labeled each post, and these labels were aggregated into a ``quality score'' $\qqq'$, the proportion of ``good'' labels. 
We find that the correlation of $\qqq'$ with $\qqq$ ($R^2 =0.25$) is more than double of that with $\ppp$ ($R^2 =0.12$), suggesting that $\qqq$ is a reasonable approximation of text quality.
The low correlation of $\qqq'$ with $\ppp$ also suggests that a community effect influences the value of $\ppp$.

Since the model was trained to predict the proportion of a post's fraction of up-votes $\ppp$, but only encodes text features (bigrams), the predicted proportion of up-votes $\qqq$ corresponds to the {\em quality} of the post's text.
In other words, when we compare changes in $\qqq$, these can be attributed to changes in the text, rather than to how a community perceives the user.\footnote{Even though the predicted proportion of up-votes $\qqq$ can be biased by user and community effects, this bias affects all posts equally (since the model is only trained on textual features).  In fact, we find the model error, $\ppp - \qqq$, to be uniformly distributed across all values of $\ppp$.}
Thus, this model allows us to assess the textual quality of the post $\qqq$, while the difference between the predicted and true proportion of up-votes ($\ppp-\qqq$) allows us to quantify community bias.

Using the textual regression model, we match pairs of users $(A,B)$ that contributed posts of similar quality, but that received very different evaluations: $A$'s post was positively evaluated, while $B$'s post was negatively evaluated.
This experimental design can be interpreted as selecting pairs of users that appear indistinguishable before the ``treatment'' (i.e., evaluation) event, but where one was \punished while the other \rewarded.
The goal then is to measure the effect of the treatment on the users' future behavior.
As the two users ``looked the same'' before the treatment, any change in their future behavior can be attributed to the effect of the treatment (i.e., the act of receiving a positive or a negative evaluation).

\begin{figure}[t]
  \centering
  \includegraphics[width=\linewidth]{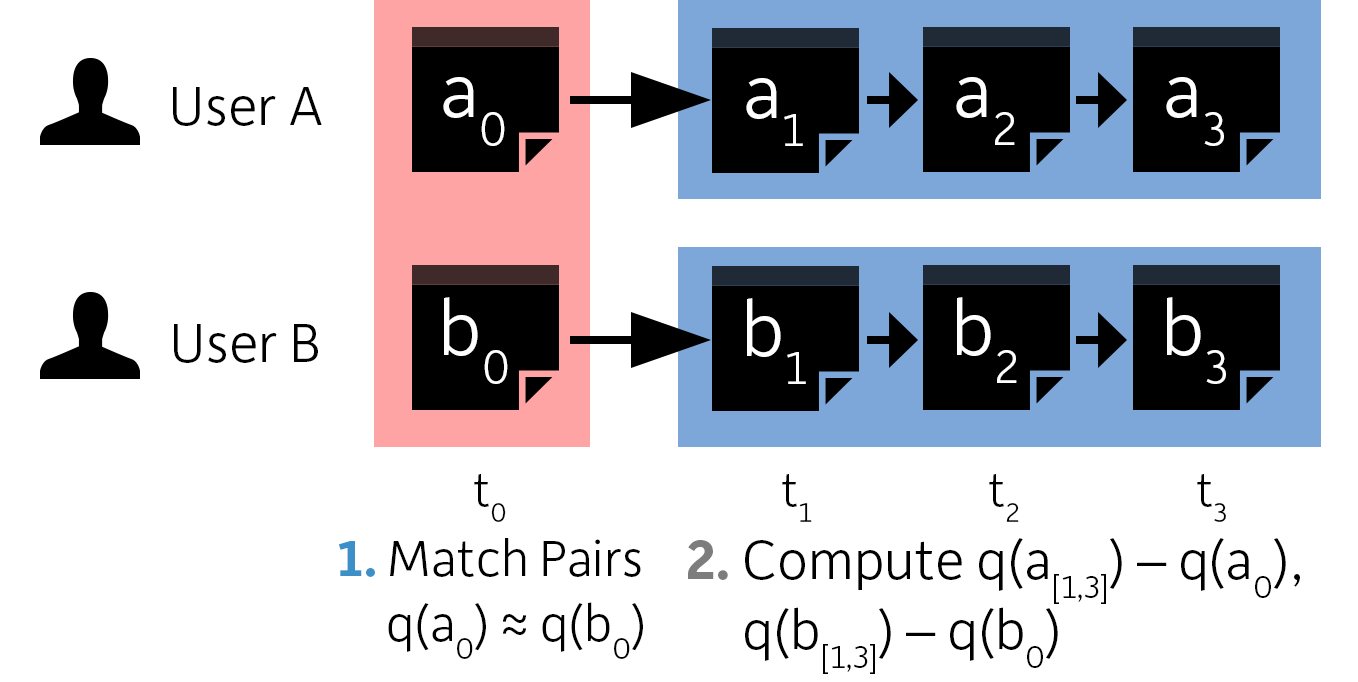}
  \caption{To measure the effects of positive and negative evaluations on post quality, we match pairs of posts of similar textual quality $\qqq(a_0)\approx\qqq(b_0)$ written by two users $A$ and $B$ with similar post histories, where $A$'s post $a_0$ received a positive evaluation, and $B$'s post $b_0$ received a negative evaluation. We then compute the change in quality in the subsequent three posts: $\qqq(a_{[1,3]})-\qqq(a_0)$ and $\qqq(b_{[1,3]})-\qqq(b_0)$.}

  \label{fig:matching_example}
\end{figure}

Figure~\ref{fig:matching_example} summarizes our experimental setting.  Here, $A$'s post $a_0$ received a positive evaluation and $B$'s post $b_0$ received a negative evaluation, and we ensure that these posts are of the same textual quality, $|\qqq(a_0) - \qqq(b_0)| \le 10^{-4}$.
We further control for the number of words written by the user, as well as for the user's past behavior: both the number of posts written before the evaluation was received, and the mean proportion of up-votes votes received on posts in the past (Table \ref{tab:matching}).
To establish the effect of \reward and \punishment we then examine the next three posts of $A$ ($a_{[1,3]}$) and the next three posts of $B$ ($b_{[1,3]}$).

It is safe to assume that when a user contributes the first post $a_1$ after being punished or rewarded, the feedback on her previous post $a_0$ has already been received. Depending on the community, in roughly 70 to 80\% of the cases the feedback on post $a_0$ at time of $a_1$'s posting is within 1\% of $a_0$'s final feedback $p(a_0)$.

\begin{table*}[!ht]
\small
\centering
\ra{1.3}
\begin{tabular}{lll|ll}\toprule
    & \multicolumn{2}{c|}{\textbf{Before matching}} & \multicolumn{2}{c}{\textbf{After matching}} \\
    & Positive ($A$) & Negative ($B$) & Positive ($A$) & Negative ($B$) \\
    & $n=72463$ & $n=39788$ & $n=35640$ & $n=35640$ \\
    \hline
    Textual quality $q(a_0)$ / $q(b_0)$ & 0.885 & 0.810 & 0.828 & 0.828 \\
    Number of words & 42.1 & 34.5 & 29.8 & 30.0 \\
    Number of past posts & 507 & 735 & 596 & 607 \\
    Prop. positive votes on past posts & 0.833 & 0.650 & 0.669 & 0.668 \\
\bottomrule
\end{tabular}
\caption{To obtain pairs of positively and negatively evaluated users that were as similar as possible, we matched these user pairs on post quality and the user's past behavior. On the CNN dataset, the mean values of these statistics were significantly closer after matching. Similar results were also obtained for other communities.}
\label{tab:matching}
\end{table*}

\subsubsection{How feedback affects a user's post quality.} To understand whether evaluations result in a change of text quality, we compare the post quality for users $A$ and $B$ before and after they receive a \punishment or a \reward.
Importantly, we do not compare the actual fraction $\ppp$ of up-votes received by the posts, but rather the fraction $\qqq$ as predicted by the text-only regression model.

By design, both $A$ and $B$ write posts of similar quality $\qqq(a_0) \approx \qqq(b_0)$
at time $t=0$. We then compute the quality of the three posts following $t=0$ as the average predicted fraction of up-votes $\qqq(a_{[1,3]})$ of posts $a_1, a_2, a_3$. Finally, we compare the post quality before/after the treatment event, by computing the difference $\Delta_a = \qqq(a_{[1,3]})-\qqq(a_0)$ for the rewarded user $A$. Similarly, we compute $\Delta_b = \qqq(b_{[1,3]}) - \qqq(b_0)$ for the punished user $B$.

Now, if the positive (respectively negative) feedback has no effect and the post quality does not change, then the difference $\Delta_a$ (respectively $\Delta_b$) should be close to zero. However, if subsequent post quality changes, then this quantity should be different from zero.
Moreover, the sign of $\Delta_a$ (respectively $\Delta_b$)  gives us the direction of change: a positive value means that the post quality of positively (respectively negatively) evaluated users improves, while a negative value means that post quality drops after the evaluation.

Using a Mann-Whitney's U test, we find that across all communities, the quality of text significantly changes after the evaluation. In particular, we find that the post quality significantly {\em drops} after a negative evaluation ($\Delta_b < 0$ at significance level $p < 0.05$ and effect size $r > 0.06$).
This effect is similar both within and across threads (average $r=0.19, 0.18$ respectively).
While the effect of negative feedback is consistent across all communities, the effect of positive feedback is inconsistent and not significant.

These results are interesting as they establish the effect of reward and punishment on the quality of a user's future posts. Surprisingly, our findings are in a sense exactly the opposite than what we would expect under the operant conditioning framework. 
Rather than evaluations increasing the user's post quality and steering the community towards higher quality discussions, we find that negative evaluations actually decrease post quality, with no clear trend for positive evaluations having an effect either way.

\subsubsection{How feedback affects community perception.}
We also aim to quantify whether evaluations changes the community's perception of the evaluated user (community bias). That is, do users that generally contribute good posts ``undeservedly'' receive more up-votes for posts that may actually not be that good? And similarly, do users that tend to contribute bad posts receive more down-votes even for posts that are in fact good? 

To measure the community perception effect we use the experimental setup already illustrated in Figure~\ref{fig:matching_example}.
As before, we first match users $(A, B)$ on the predicted fraction of up-votes $\qqq$; we then measure the residual difference between the {\em true} and the {\em predicted} fraction of up-votes $\ppp(a_{[1:3]}) - \qqq(a_{[1:3]})$ after user $A$'s treatment (analogously for user $B$). Systematic non-zero residual differences are suggestive of community bias effects, i.e., posts get evaluated differently from how they should be based solely on their textual quality. Specifically, if the community evaluates a user's posts higher than expected then the residual difference is positive, and if a user's posts are evaluated lower than expected then the residual difference is negative.


Across all communities, posts written by a user after receiving negative evaluations are perceived worse than the text-only model prediction, and this discrepancy is much larger than the one observed after positive evaluations ($p < 10^{-16}, r > 0.03$). This effect is also stronger within threads (average $r = 0.57$) than across threads (average $r = 0.13$). For instance,  after a negative evaluation on CNN, posts written by the punished author in the same thread are evaluated on average 48 percentage points lower than expected by just considering their text.

Note that the true magnitude of these effects could be smaller than reported, as using a different set of textual features could result in a more accurate classifier, and hence smaller residuals. Nevertheless, the experiment on post quality presented earlier does not suffer from these potential classifier deficiencies.

\begin{figure}[t]
  \centering
  \includegraphics[width=\linewidth]{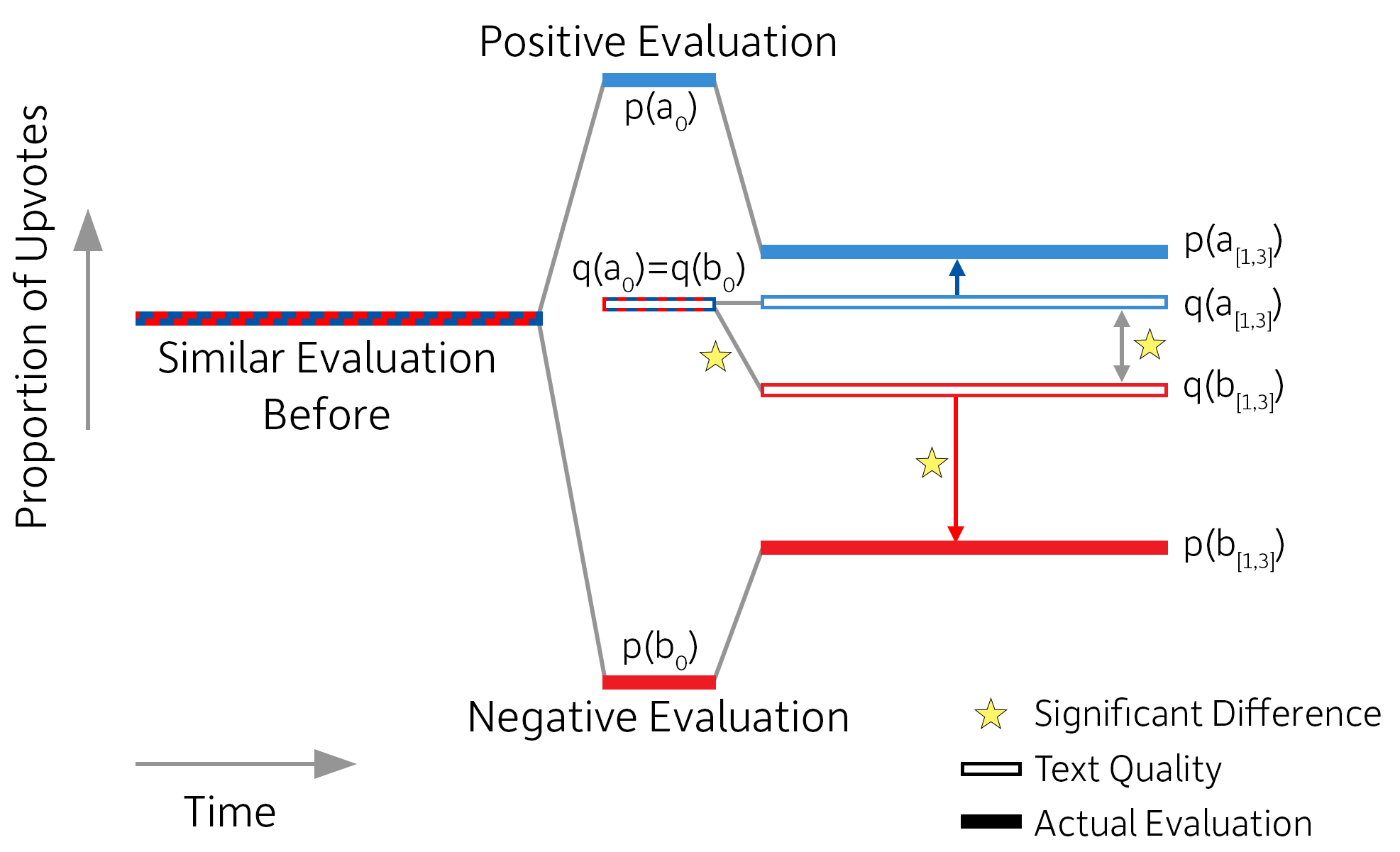}
  \caption{The effect of evaluations on user behavior. We observe that both community perception and text quality is significantly worse after a negative evaluation than after a positive evaluation (in spite of the initial post and user matching). Significant differences are indicated with stars, and the scale of effects have been edited for visibility. 
  }
  \label{fig:perception_and_quality}
  \vspace{-3mm}
\end{figure}

\subsubsection{Summary.}
Figure~\ref{fig:perception_and_quality} summarizes our observations regarding the effects of community feedback on the textual and perceived quality of the author's future posts. We plot the textual quality and proportion of up-votes before and after the evaluation event (``treatment''). Before the evaluation the textual quality of the posts of two users $A$ and $B$ is indistinguishable, i.e., $q(a_0) \approx q(b_0)$. However, after the evaluation event, the textual quality of the posts of the positively evaluated user $A$ remains at the same level, i.e., $q(a_0)\approx q(a_{[1:3]})$, while the quality of the posts of the negatively evaluated user $B$ drops significantly, i.e., $q(b_0) > q(b_{[1:3]})$.  We conclude that community feedback does not improve the quality of discussions, as predicted by the operand conditioning theory. Instead, \punished authors actually write worse in subsequent posts, while \rewarded authors do not improve significantly.

We also find suggestive evidence of community bias effects, creating a discrepancy between the perceived quality of a user's posts and their textual quality.  This perception bias appears to mostly affect negatively evaluated users: the perceived quality of their subsequent posts $p(b_{[1:3]})$ is much lower than their textual quality $q(b_{[1:3]})$, as illustrated in Figure~\ref{fig:perception_and_quality}.  Perhaps surprisingly, we find that community perception is an important factor in determining the proportion of up-votes a post receives. 

Overall, we notice an important asymmetry between the effects of positive and negative feedback: the detrimental effects of punishments are much more noticeable than the beneficial effects of rewards.  This asymmetry echoes the {\em negativity effect} studied extensively in the social psychology literature~\cite{Kanouse:NegativityInEvaluations:1972,baumeister2001bad}.



\section{User Activity}
\label{sec:user}

Despite the detrimental effect of community feedback on an author's content quality, community feedback could still have a beneficial effect by selectively regulating {\em quantity}, i.e., discouraging contributions from \punished authors and encouraging \rewarded authors to contribute more. 

To establish whether this is indeed the case we again use a methodology based on propensity score matching, where our variable of interest is now posting frequency. As before, we pair users that wrote posts of the same textual quality (according to the textual regression model), ensuring that one post was positively evaluated, and the other negatively evaluated.
We further control for the variable of interest by considering matching pairs of users that had the same posting frequency before the evaluation.
This methodology allows us to compare the effect of positive and negative feedback on the author's future posting frequency.

\begin{figure}
  \centering
  \includegraphics[width=1\linewidth]{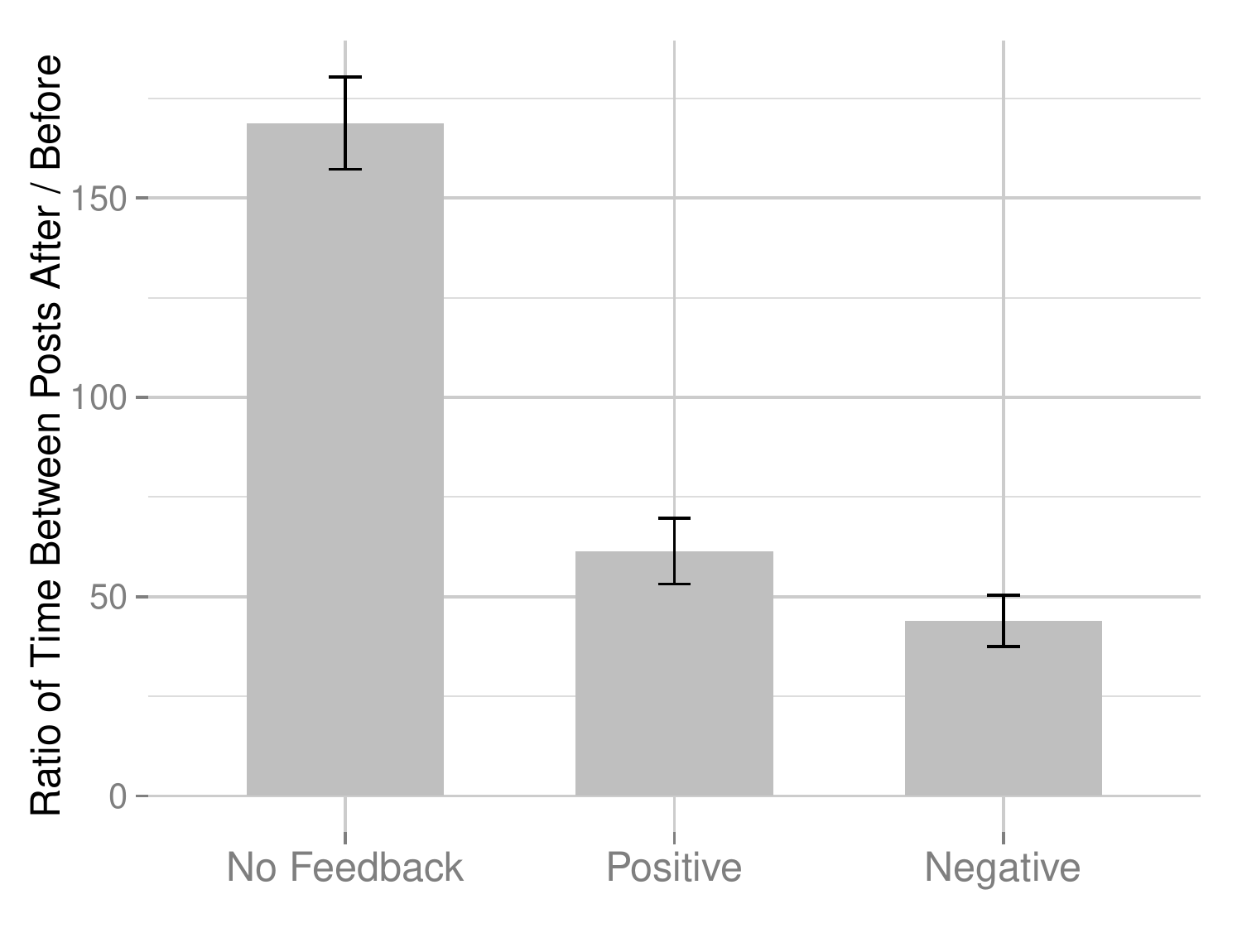}
  \vspace{-1.5em}
  \caption{Negative evaluations increase posting frequency more than positive evaluations; in contrast, users that received no feedback slow down. (Values below 100 correspond to an increase in posting frequency after the evaluation; a lower value corresponds to a larger increase.)}
  \label{fig:frequency}
  \vspace{-0.5em}
\end{figure}

Figure \ref{fig:frequency} plots the ratio between inverse posting frequency after the treatment and inverse posting frequency before the treatment, where inverse frequency is measured as the average time between posts in a window of three posts after/before the treatment. Contrary to what operant conditioning would predict, we find that negative evaluations encourage users to post more frequently. Comparing the change in frequency of the \punished users with that of the \rewarded users, we also see that negative evaluations have a greater effect than positive evaluations ($p < 10^{-15}, r > 0.18$).
Moreover, when we examine the users who received no feedback on their posts, we find that they actually slow down. In particular, users who received no feedback write about 15\% less frequently, while those who received positive feedback write 20\% more frequently than before, and those who received negative feedback write 30\% more frequently than before. These effects are also statistically significant, and consistent across all four communities.

\begin{figure}[t]
  \centering
  \includegraphics[width=\linewidth]{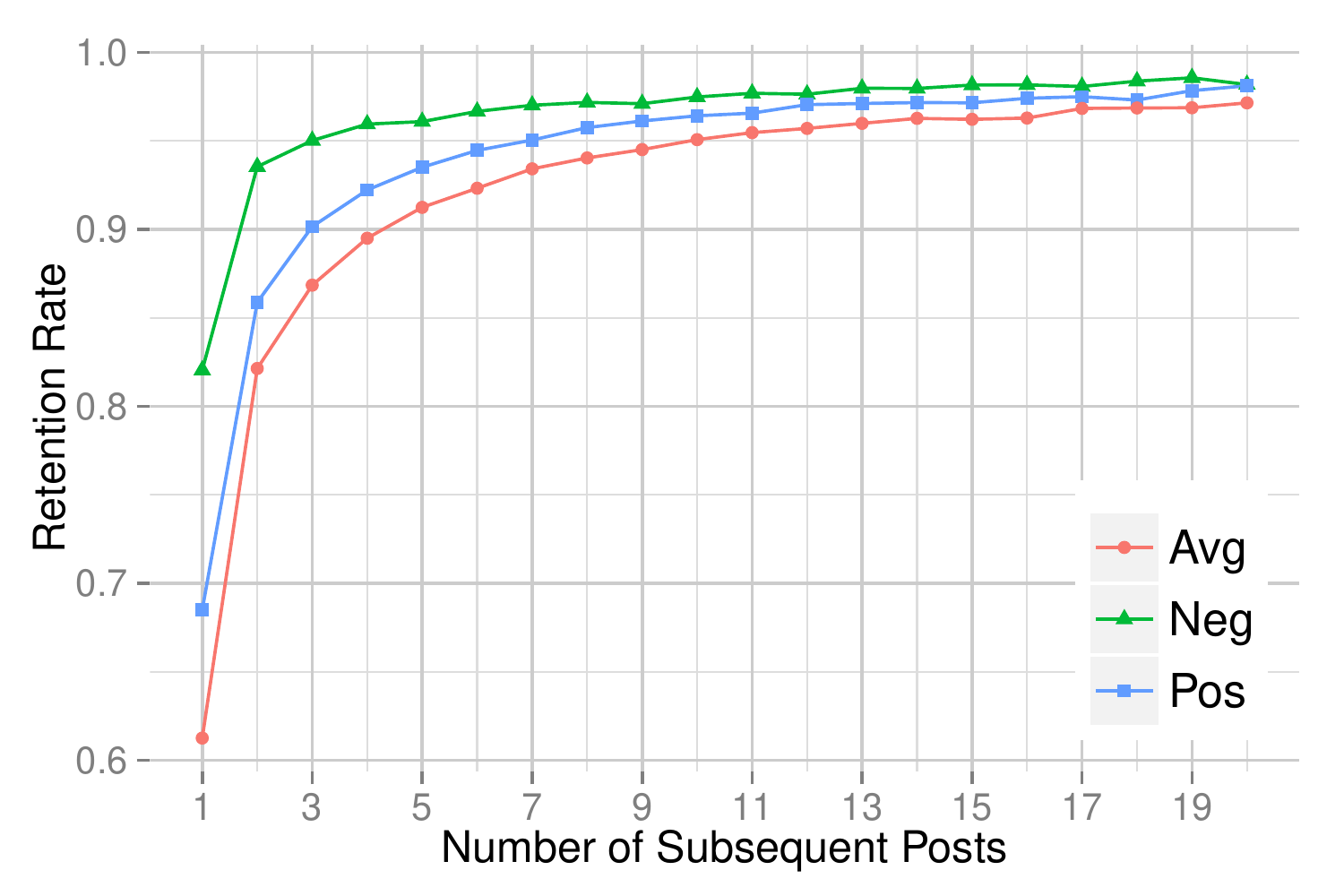}
  \caption{Rewarded users (``Pos'') are likely to leave the community sooner than punished users (``Neg''). Average users (``Avg'') are most likely to leave. For a given number of subsequent posts $x$ the retention rate is calculated as the fraction of users that posted at least $x$ more posts.
  }
  \label{fig:orig_retention}
  \vspace{-3mm}
\end{figure}

The same general trend is true when considering the impact of evaluations on user retention (Figure \ref{fig:orig_retention}): \punished users (``Neg'') are more likely than  \rewarded users (``Pos'') to stay in the community and contribute more posts ($\chi^2 > 6.8, p < 0.01$); also both types or users are less likely to leave the community than the control group (``Avg'').
Note, however, that the nature of this experiment does not allow one to control for the value of interest (retention rate) before the evaluation.

The fact that both types of evaluations encourage users to post more frequently suggests that providing negative feedback to ``bad'' users might not be a good strategy for combating undesired behavior in a community. Given that users who receive no feedback post less frequently, a potentially effective strategy could be to ignore undesired behavior and provide no feedback at all.

\section{Voting Behavior}
\label{sec:enemies}

Our findings so far suggest that negative feedback worsens the quality of future interactions in the community as  \punished users post more frequently.
As we will discuss next, these detrimental effects are exacerbated by the changes in the voting behavior of evaluated users.

\subsubsection{Tit-for-tat.} 
\begin{figure}[t]
  \centering
  \includegraphics[width=\linewidth]{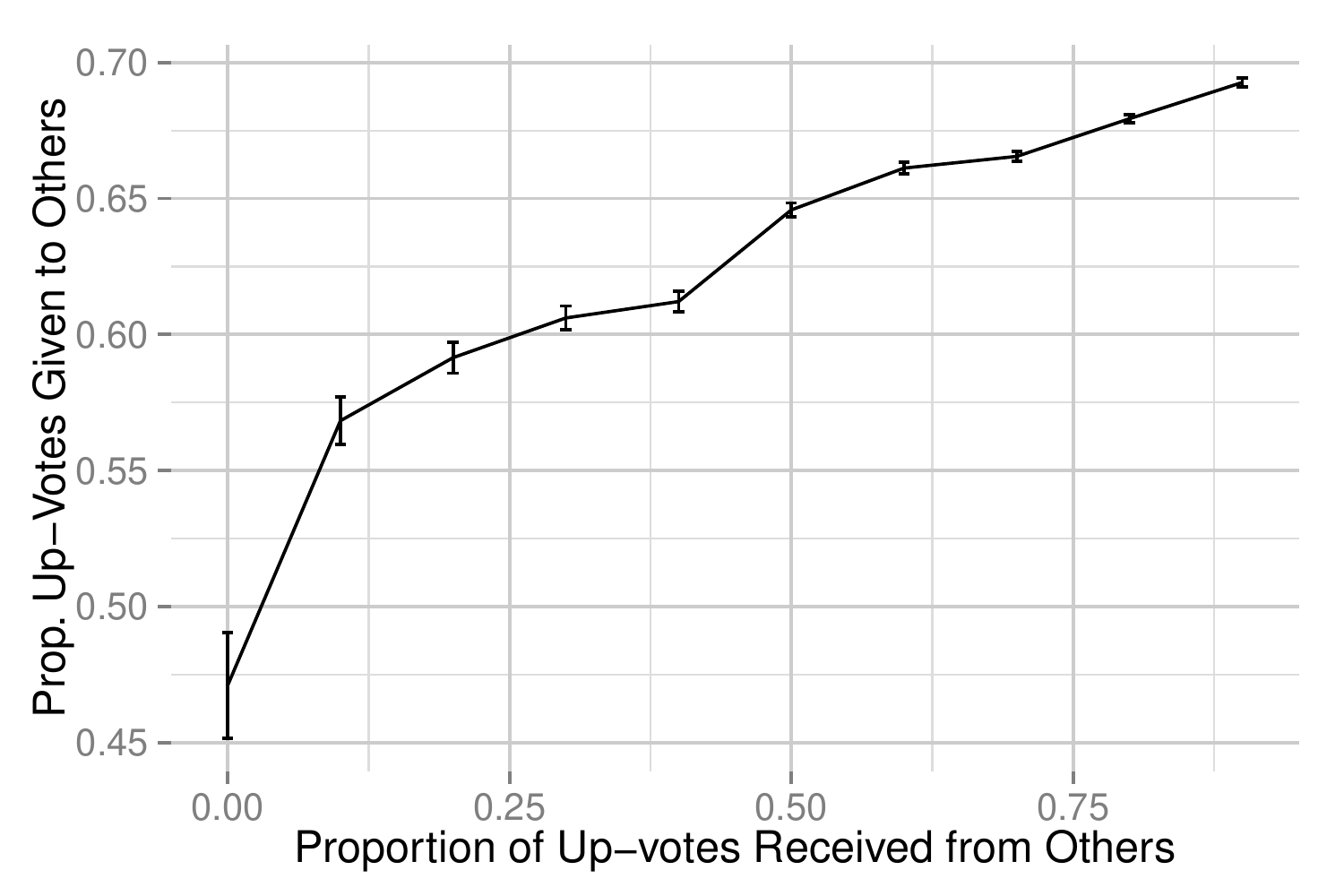}
  \caption{Users seem to engage in ``tit-for-tat'' --- 
  the more up-votes a user receives, the more likely she is to give up-votes to content written by others.}
  \label{fig:give_receive}
\end{figure}

As users receive feedback, both their posting and voting behavior is affected. When comparing the fraction of up-votes received by a user with the fraction of up-votes given by a user, we find a strong linear correlation (Figure~\ref{fig:give_receive}).
This suggests that user behavior is largely ``tit-for-tat''.
If a user is negatively/positively evaluated, she in turn will negatively/positively evaluate others.
However, we also note an interesting deviation from the general trend.
In particular, very negatively evaluated people actually respond in a positive direction: the proportion of up-votes they give is {\em higher} than the proportion of up-votes they receive.
On the other hand, users receiving many up-votes appear to be more ``critical'', as they evaluate others more negatively.
For example, people receiving a fraction of up-votes of 75\% tend to give up-votes only 67\% of the time.

\begin{figure}[t]
  \centering
  \includegraphics[width=\linewidth]{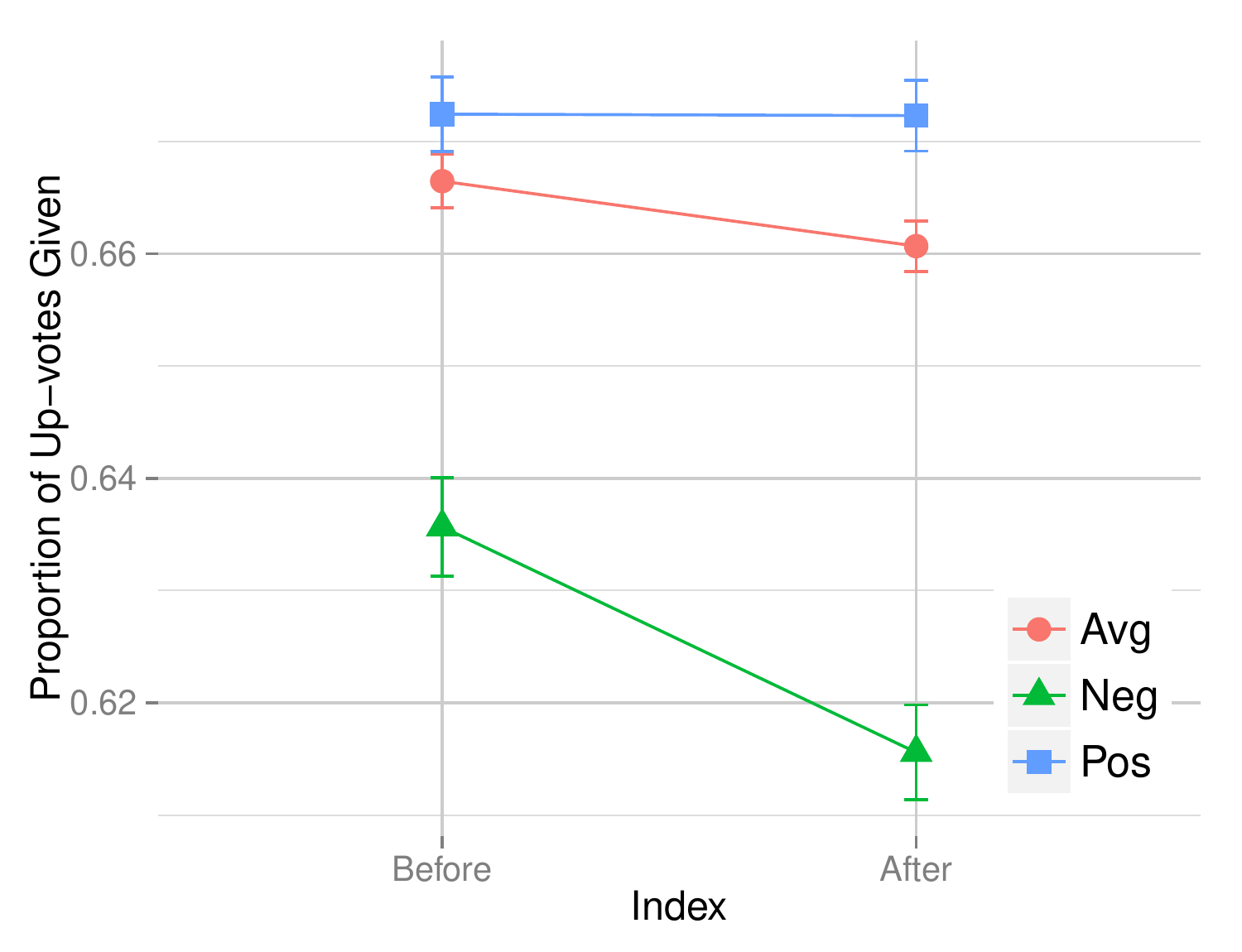}
  \caption{If a user is evaluated negatively, then she also tends to vote on others more negatively in the week following the evaluation than in the week before the evaluation. However, we observe no statistically significant effect on users who receive a positive evaluation.
  }
  \label{fig:votes_given_prop_d_base}
\end{figure}

Nevertheless, this overall perspective does not directly distinguish between the effects of positive and negative evaluations on voting behavior.
To achieve that, Figure \ref{fig:votes_given_prop_d_base} compares the change in voting behavior following a positive or negative evaluation. 
We find that \emph{negatively-evaluated users are more likely to down-vote others} in the week following an evaluation, than in the week before it ($p < 10^{-13}, r > 0.23$).
In contrast, we observe no significant effect for the positively evaluated users.

Overall, \punished users not only change their posting behavior, but also their voting behavior by becoming more likely to evaluate their fellow users negatively.  Such behavior can percolate the detrimental effects of negative feedback through the  community.

\section{Organization of Voting Networks}
\label{sec:organization}

\begin{figure}[t]
  \centering
  \includegraphics[width=\linewidth]{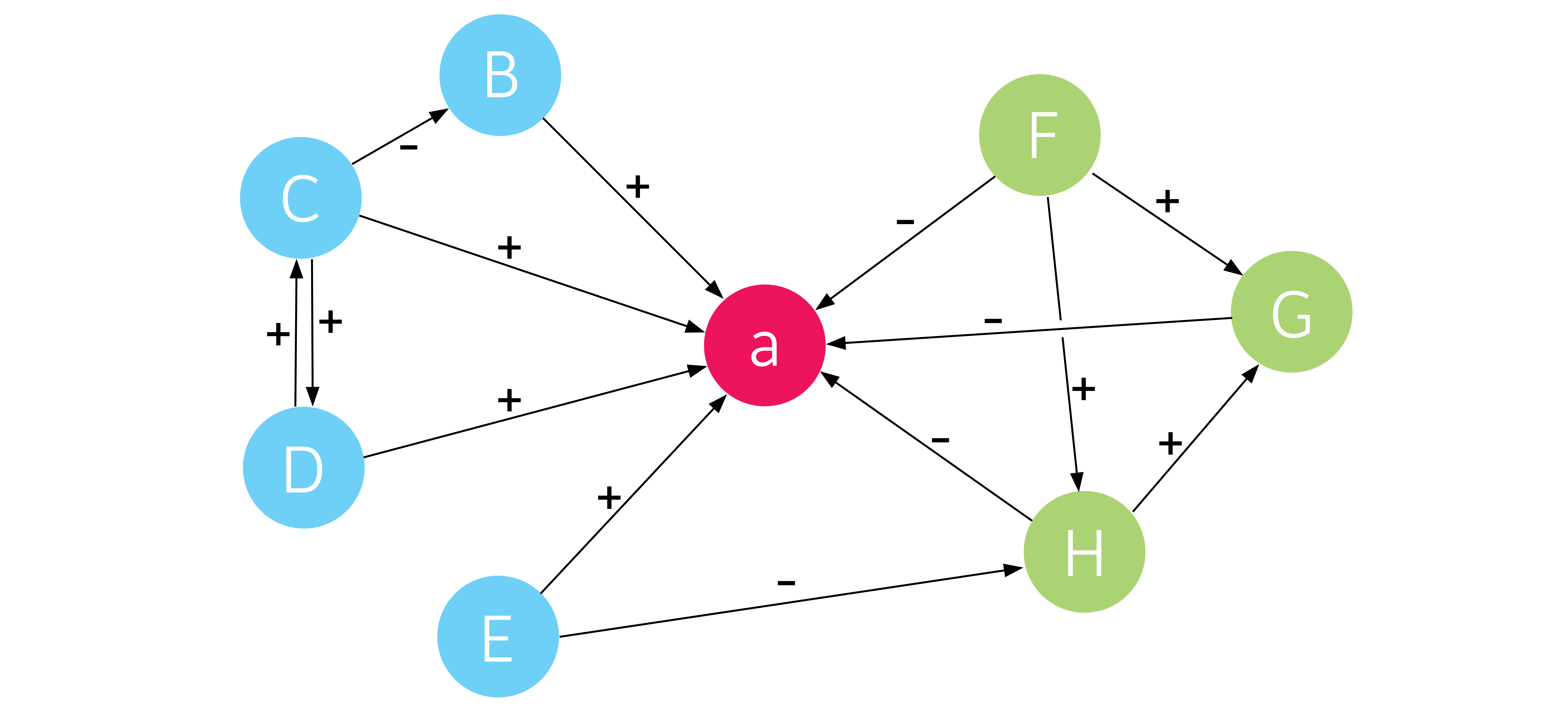}
  \caption{An example voting network $G$ around a post $a$. Users $B$ to $H$ up-vote (+) or down-vote (-) $a$, and may  also have voted on each other. The graph induced on the four users who up-voted $a$ ($B, C, D, E$) forms $G_+$ (blue nodes), and that induced on the three users who down-voted $a$ ($F, G, H$) forms $G_-$ (green nodes).}
  \label{fig:balance_example}
  \vspace{-3mm}
\end{figure}

Having observed the effects community feedback has on user behavior, we now turn our attention to structural signatures of positive and negative feedback.
In particular, we aim at studying the structure of the social network around a post and understanding \begin{inparaenum}[(1)] \item when do evaluations most polarize this social network, and \item whether positive/negative feedback comes from independent people or from tight groups.\end{inparaenum}

\subsubsection{Experimental setup.}
We define a social networks around each post, a {\em voting network}, as illustrated in Figure~\ref{fig:balance_example}.
For a given post $a$, we generate a graph $G = (V,E)$, with $V$ being the set of users who voted on $a$.
An edge $(B,C)$ exists between voters $B$ and $C$ if $B$ voted on $C$ in the 30 days prior to when the post $a$ was created.
Edges are signed: positive for up-votes, negative for down-votes.
We examine voting networks for posts which obtained at least ten votes, and have at least one up-vote and one down-vote. 

\subsubsection{When is the voting network most polarized?} 
And, to what degree do coalitions or factions form in a post's voting network?
As our networks are signed, we apply structural balance theory \cite{cartwright1956structural}, and examine the fraction of balanced triads in our network.
A triangle is balanced if it contains three positive edges (a set of three ``friends''), or two negative edges and a positive edge (a pair of ``friends'' with a common ``enemy''). The more balanced the network, the stronger the separation of the network into coalitions --- nodes inside the coalition up-vote each other, and down-vote the rest of the network.

\begin{figure}[t]
  \centering
  \subfloat[][Balance]{\includegraphics[width=0.50\linewidth]{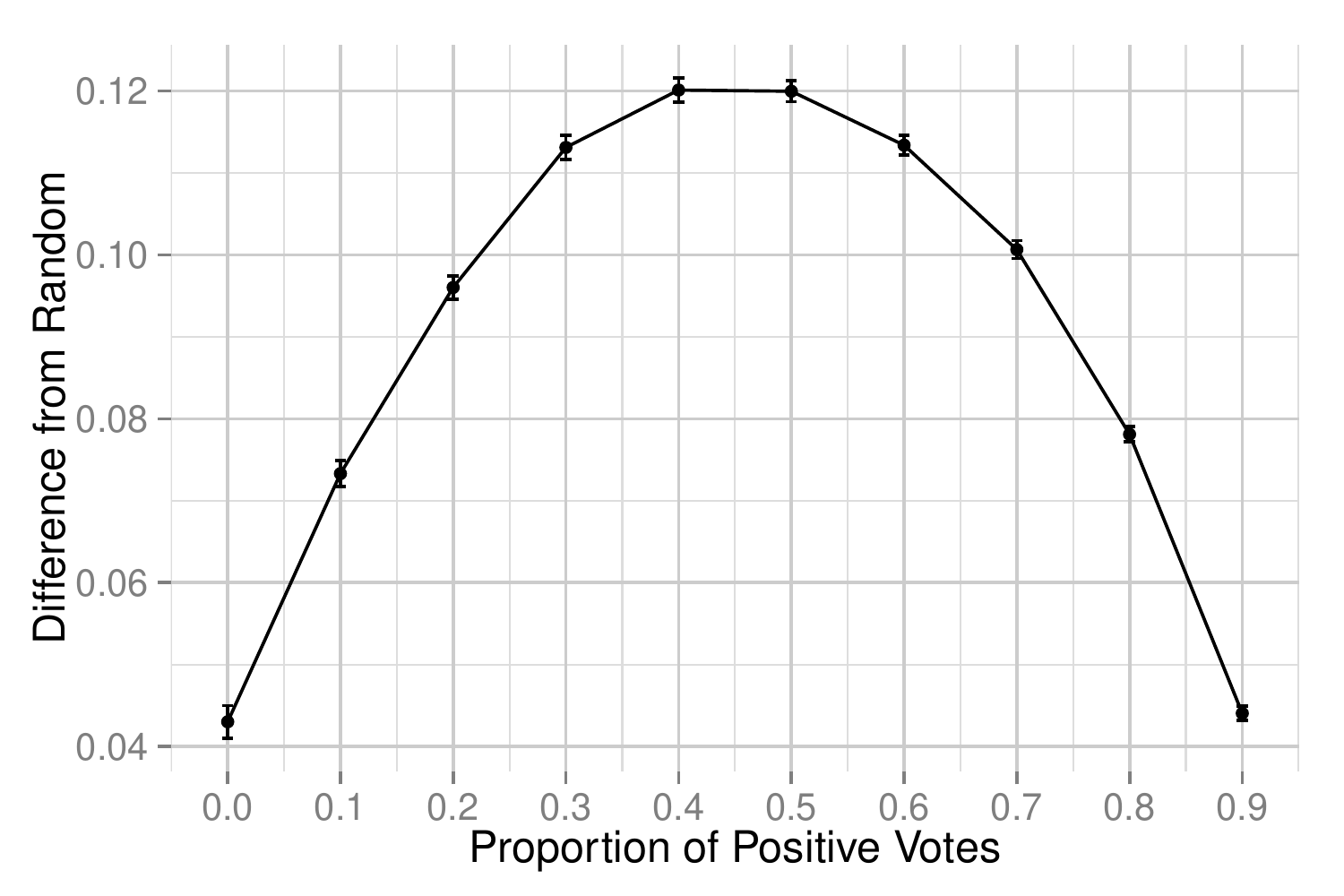}\label{fig:graph_balance}}
  \subfloat[][Edges across the camps]{\includegraphics[width=0.50\linewidth]{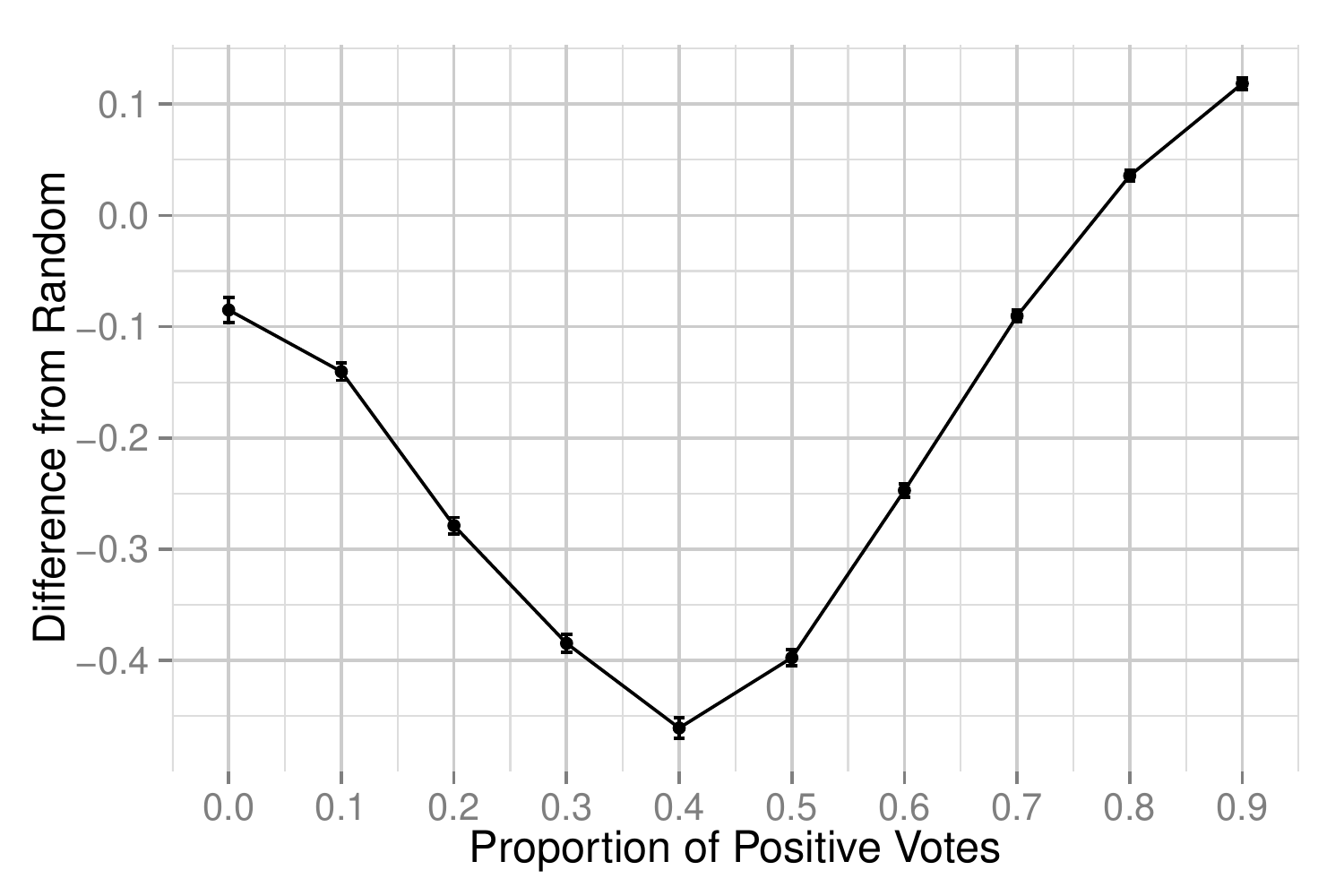}\label{fig:graph_edge}}
  \caption{(a) The difference between the observed fraction of balanced triangles and that obtained when edge signs are shuffled at random. The peak at 50\% suggests that when votes are split evenly, these voters belong to different groups. (b) The difference of the observed number of edges between the up-voters and negative voters, versus those in the case of random rewiring. The lowest point occurs when the votes are split evenly.}
\end{figure}

Figure~\ref{fig:graph_balance} plots the fraction of balanced triangles in a post's voting network as a function of the proportion of up-votes that post receives (normalized by randomly shuffling the edge signs in the original voting network).
When votes on a post are split about evenly between up- and down-votes, the network is most balanced.
This means that when votes are split evenly, the coalitions in the network are most pronounced and thus the network most polarized.
This observation holds in all four studied communities.

We also compare the number of edges between the up-voters and down-voters (i.e., edges crossing between $G_+$ and $G_-$) to that of a randomly rewired network (a random network with the same degree distribution as the original).
Figure \ref{fig:graph_edge} plots the normalized number edges between the two camps as a function of the proportion of up-votes the post received. We make two interesting observations.
First, the number of edges crossing the positive and negative camp is lowest when votes are split about evenly.
Thus, when votes are split evenly not only is the network most balanced, but also the number of edges crossing the camps is smallest.
Second, the number of edges is always below what occurs in randomly rewired networks (i.e., the Z-scores are negative).
This suggests that the camp of up-voters and the camp of down-voters are generally not voting on each other.
These effects are qualitatively similar in all four communities.

\subsubsection{Where does feedback come from?}
Having observed the formation of coalitions, we are next interested in their relative size.
Is feedback generally given by isolated individuals, or by tight groups of like-minded users?
We find interesting differences between communities. In general-interest news sites like CNN, up-votes on positively-evaluated posts are likely to come from multiple groups --- the size of the largest connected component decreases as the proportion of up-votes increases.
In other words, negative voters on a post are likely to have voted on each other.
However, on special-interest web sites like Breitbart, IGN, and Allkpop, the size of the largest connected component also peaks when votes are almost all positive. 
Thus, up-voters in these communities are also likely to have voted on each other suggesting that they come from tight groups.

\section{Discussion}
\label{sec:discussion}

Rating, voting and other feedback mechanisms are heavily used in today's social media systems, allowing users to express opinions about the content they are consuming.
In this paper, we contribute to the understanding of how feedback mechanisms are used in online systems, and how they affect the underlying communities.  We start from the observation that when users evaluate content contributed by a fellow user (e.g., by liking a post or voting on a comment) they also implicitly evaluate the author of that content, and that this can lead to complex social effects.

In contrast to previous work, we analyze effects of feedback at the user level, and validate our results on four large, diverse comment-based news communities.
We find that negative feedback leads to significant changes in the author's behavior, which are much more salient than the effects of positive feedback.  These effects are detrimental to the community: authors of negatively evaluated content are encouraged to post more, and their future posts are also of lower quality.
Moreover, these punished authors are more likely to later evaluate their fellow users negatively, percolating these undesired effects through the community.

We relate our empirical findings to the operand conditioning theory from behavioral psychology, which explains the underlying mechanisms behind reinforcement learning, and find that the observed behaviors deviate significantly from what the theory predicts.
There are several potential factors that could explain this deviation.
Feedback in online settings 
is potentially very different from that in controlled laboratory settings. 
For example, receiving down-votes is likely a much less severe punishment than receiving electric shocks.  Also, feedback effects might be stronger if a user trusts the authority providing feedback, e.g., site administrators down-voting author's posts could have a greater influence on the author's behavior than peer users doing the same.

Crucial to the arguments made in this paper is the ability of the machine learning regression model to estimate the textual quality of a post.
Estimating text quality is a very hard machine learning problem, and although we validate our model of text quality by comparing its output with human labels, the goodness of fit we obtain can be further improved.
Improving the model could allow for finer-grained analysis and reveal even subtler relations between community feedback and post quality.

Localized back-and-forth arguments between people (i.e. flame wars) could also potentially affect our results. However, we discard these as being the sole explanation since the observed behavioral changes carry on across different threads, in contrast to flame wars which are usually contained within threads. 
Moreover, we find that the set of users providing feedback changes drastically across different posts of the same user.
This suggests that users do not usually create ``enemies'' that continue to follow them across threads and down-vote any posts they write.  Future work is needed to  understand the scale and effects of such behavior.

There are many interesting directions for future research.
While we focused only on one type of feedback --- votes coming from peer users --- there are several other types that would be interesting to consider, such as feedback provided through textual comments.
Comparing voting communities that support both up- and down-votes with those that only allow for up-votes (or likes) may also reveal subtle differences in user behavior.
Another open question is how the relative authority of the feedback provider affects author's response and change in behavior.
Further, building machine learning models that could identify which types of users improve after receiving feedback and which types worsen could allow for targeted intervention.
Also, we have mostly ignored the content of the discussion, as well as the context in which the post appears.
Performing deeper linguistic analysis, and understanding the role of the context may reveal more complex interactions that occur in online communities.

More broadly, online feedback also relates to the key sociological issues of norm enforcement and socialization \cite[inter alia]{Parsons:TheSocialSystem:1951,Cialdini:SocialInfluenceSocialNormsConformityAndCompliance:1998}, i.e. what role does peer feedback play in directing users towards the behavior that the community expects, and how a user's reaction to feedback can be interpreted as her desire to conform to (or depart from) such community norms.
For example, our results suggest that negative feedback could be a potential trigger of deviant behavior (i.e., behavior that goes against established norms, such as trolling in online communities).
Here, surveys and controlled experiments can complement our existing data-driven methodology and shed light on these issues.

\paragraph{Acknowledgments} We thank Disqus for the data used in our experiments and our anonymous reviewers for their helpful comments.
This work has been supported in part by a Stanford Graduate Fellowship, an Alfred P. Sloan Fellowship, a Microsoft Faculty Fellowship and NSF IIS-1159679.

\bibliographystyle{aaai}
\small
\bibliography{refs}

\end{document}